\documentclass{IOS-Book-Article}

\usepackage{amsmath}
\usepackage{paralist}
\usepackage{graphicx}
\usepackage{tabularx}
\usepackage{soul, color}
\sethlcolor{green}
\usepackage{eurosym}
\usepackage{url}

\newtheorem{theorem1}{Theorem}
\newtheorem{definition}[theorem1]{Definition}
  \newtheorem{example}{\textbf{Example}}

\usepackage{mathptmx}

\def\hb{\hbox to 10.7 cm{}}

\begin{document}

\pagestyle{headings}
\def\thepage{}

\begin{frontmatter}              

\title{Why a computer program is a functional whole\thanks{This technical report is a substantially longer version of the paper entitled ``The computer program as a functional whole'' accepted at FOIS2020. This report mainly has a much longer introduction, more examples, and at times informal writing style in an attempt to broaden accessibility of the content to a larger audience.}}


\author{\fnms{C. Maria} \snm{KEET}%
}

\runningauthor{C. M. Keet}
\address{Department of Computer Science, University of Cape Town, South Africa\\ mkeet@cs.uct.ac.za}

\begin{abstract}
Sharing, downloading, and reusing software is common-place, some of which is carried out legally with open source software. When it is not legal, it is unclear just how many copyright  infringements and trade secret violations have taken place: does an infringement count for the artefact as a whole or perhaps 
for each file of the program? 
To answer this question, it must first be established whether a program should be considered as an integral whole, a collection, or a mere set of distinct files, and why.
We argue that a program is a functional whole, availing of, and combining, arguments from mereology, granularity, modularity, unity, and function to substantiate the claim. The argumentation and answer contributes to the ontology of software artefacts, may assist industry in litigation cases, and demonstrates that the notion of unifying relation is operationalisable. Indirectly, it provides support for continued modular design of artefacts following established engineering practices. 
\end{abstract}

\end{frontmatter}
\markboth{July 2020\hb}{July 2020\hb}

\section{Introduction}
\label{sec:intro}

End users of Information Technology are familiar with software `apps', i.e., computer programs, that can be launched through a one or double mouse-click or a finger-tap action on the screen of the device. From a user experience perspective, the computer program may appear to be a single thing. 
There are typically multiple files involved in the running of a computer program for it to carry out its principal function\footnote{Note: software, computer program, (software) application, and (software) tool are often used interchangeably, but are not all one and the same thing. A software application or tool, such as Adobe Acrobat Reader, is a particular type of computer program, since it is a computer program that needs an operating system to be able to run and perform its function. Software is a broader term and also includes other files on the computer (including its external drives).} 
however, due to the increasingly complex software nowadays, as well as the move toward modularisation in software design and development over at least the past 25 years. 
For instance, compare an Wordperfect v3.x word processor with MS Word with what-you-see-is-what-you-get text processing, mail merge, and inserting tables, or texting an SMS compared to all the WhatsApp features that has a range of facilitating sub-functions such as encryption, saving conversations, and reporting status updates. `Under the hood' outside of the end user's immediate view, there are typically multiple files to make such apps work. It is not impossible that there is only one single file as source code and as executable, but is is very rare and such a tool is typically not for regular consumption as industry-grade software. For instance, a {\tt hello.c} source code file to print just ``Hello World'' on screen still calls a built-in {\tt print} function that is stored in another file than the typed-up code.

A computer program may exist in different `formats': human-readable source code that is normally written by humans or as compiled or interpreted code (which is then eventually object (machine) code) that is the executable that the machine will use. The program typically uses components internally and any component may `call' one or more other components (files) for its proper functioning. This is often hidden from the user view, since it is irrelevant for the user how exactly the program achieves what it is supposed to achieve, just like one isn't subjected to the inner workings of a car when driving the car. Such component files are stored in some specific directory and are artefacts---i.e., human-made things---such as plugins for a larger program, dynamic linked libraries (dll's), icons for the interfaces and so forth, as well as configuration files to, e.g., set the language of the program or how much computer resources it is allowed to use. Disregarding those finer-grained details, one still may consider, e.g., MS Word, Google Chrome, or Tinder, or their respective source code, a single artefact that is downloaded and installed. If you `pirate', say, McAfee Antivirus software, Microsoft Office, or the Adobe Suite, 
you presumably would break the law once for each program, not hundreds of times, once for each component file in the installation. 
Why is this so? Or, conversely: could one possibly be subjected to be sued for many infringements if the program might turn out {\em not} be a whole upon closer inspection, as if it were a mere set of independent objects, and each file is going to cost you money or time in prison? 
Why may one care about the answer?

Aside from intellectual inquiry, there are many practical and financial consequences, to the point that there are, and have been, multiple court cases on copyright infringements, trade secret violations, and patent misuse of programs. For illustrative purpose of such type of litigations, one could consider internationally widely debated ones, such as the ongoing Google vs Oracle case on whether a programming interface can be copyrighted or not \cite{Samuelson19} and a recently settled case between Cisco and Arista on computer network management tools where Arista eventually payed Cisco \$ 400 million \cite{Brachmann2018}, 
but there are also ongoing local (South African) cases, such as about software for medical schemes \cite{Cairns18}. 
Precedent-setting ones involve, among others,  
the notion of non-literal similarity of the source code and the, as it turned out, free use of copyrighted text for the purpose of training an algorithm. Closely related issues include the `patent wars' on software patents and their infringements and whether data can be copyrighted or not.

One of the issues some of these litigation cases raise, amounts to a question of an ontological nature, which is introduced here in an general way and with broader applicability. One party claims that the computer program is not one whole, but that for the purpose of litigation, there are the, say, $n=1000$,  source files that allegedly have been copied and the claimant wants the defendant to be fined for the 1000 copyright violations, once for each source file; hence, not, say, a \euro100 fine for the one infringement of the program, but a fine of \euro100000 altogether. The defendant, for obvious reasons, would rather prefer to pay the \euro100 for one infringement, if infringement were to be deemed to be the case according to the court. Such an argument can likewise be constructed for trade secret violations for stealing the intellectual property of an app or even a whole operating system, for pirated software that was illegally copied, downloading textbooks in website format (which has multiple files cf. the `one pdf' option) or a single Game of Thrones episode that happens to have as components, 
say, one audio file, 
one video file, and one 
subtitle file.

On the surface, this may sound similar to the well-known case in philosophy on the collapsing bridge. However, that typical example is about one event vs. multiple collapsing events, like many small cracks before it finally collapses and drawing event boundaries. Here, to stay with the analogy of a bridge, the `collapsing bridge' argument is rather one of ``is the number of collapses equal to the number of the parts of the bridge?''. Clearly, for the bridge, there is one collapsing event regardless how many parts the engineer used to build the bridge. If negligence has been found in the actual construction, then there is one fine for the culpability of bad bridge-building by the responsible construction company, not thousands of fines, one for each screw, cable, metal pole, and brick that the bridge was made of. 
Similarly, if someone steals a car, then there is one culpable act, not thousands for each part of the car. 
One easily can draw a parallel with computer programs: they are functional wholes, and any unlawful act, such as illegal downloads or copying copyrighted source code of a program (not under a public licence, such as GPL), 
tally up for the wholes, not for its, possibly thousands, of constituent parts.

While this reasoning by analogy may already suffice as argument to some, one could dig deeper into computer programs to assess the strength of the analogy. This means that it first needs to be established how all those files of a computer program relate, and of the source code in particular since that may be subject to copyright in some, but not all, cases. This may not seem as straightforward as the structural components of a bridge or a car. 
The main question to answer, then is: 
\begin{enumerate}
	\item[Q1:] Is a computer program (or source code) a) a (tight) {\em whole} with {\em parts}, b) a whole that is a {\em collection} of {\em artefacts}, or c) just a {\em set} of {\em artefacts} where each element is a separate self-standing, independent, item?
\end{enumerate}
This question generates two more specific ones to answer regarding the parts, in order to be able to answer Q1:
\begin{enumerate}	
	\item[Q2:] What is the relation between the files of a computer program (resp. source code) and the computer program (resp. source code)?
	\item[Q3:] What is the relation among the files of a computer program (resp. source code)?	
\end{enumerate}
An answer as to the ontology of the computer program as a whole may also elucidate what exactly copyright infringement really means for a computer program and, specifically, help draw boundaries of what is, and what is not, part of the program.  

There is ample documentation with `just so' statements about the compositionality of software, i.e., that a program consists of parts that make up the whole, 
and why that is essential to good software design practices (e.g., \cite{Bjorner10,Noback18,Prehofer08}) and to system design more generally \cite{Tripakis16}. Yet, to the best of our knowledge, there is no extant argumentation regarding the wholeness of a computer program or what it is that makes it a whole consisting of parts, why, and how from an ontological perspective.  
The more common question---that is investigated in philosophy in particular---is the nature of a computer program, like whether it is a process or of some other category \cite{Turner19} or when an artefact can be classified as being a computer program. 
They are considerations {\em at that level of granularity} of the artefact, not about its compositional nature. Such arguments do, however, strengthen the argument that a computer program is a whole. 

To answer the questions, we take insights from part-whole relations and mereology, the notion of unity, and basics of granularity, to combine them with modes of participation and functional parthood. Computer programs have a main function that is carried out by genuine functional parts and those parts have a specific unifying relation among each other at that finer-grained level of granularity, where one implies the other, therewith making a computer program a functional whole.

In the remainder of the paper, we first introduce some preliminaries in Section~\ref{sec:prelim}. We then consider the `vertical' relation between the whole and its components in Section~\ref{sec:vertical}, the `horizontal' relation between the components in Section~\ref{sec:horizontal}, and close the argument that software is a proper (complex) whole in Section~\ref{sec:whole}. We conclude in Section~\ref{sec:concl}.

\section{Preliminaries}
\label{sec:prelim}

There are many aspects to computer programs, but only a subset is needed to answer the questions posed in Section~\ref{sec:intro}, as it concerns the notion of compositionally specifically, which can be answered with going into a debate about, say, what copies of files are or whether a program is a static ting or a running process\footnote{In more detail, out of scope are topics such as whether a computer program is a process, a disposition, or an endurant, or of another main category typically found in a foundational ontology (i.e., the ontological nature of a computer program), 
 the nature and multiplication aspects of so-called `information objects', 
and digital copies with their identity, such as that the files stored on disk temporarily reside in main memory.}. 
The argumentation for it avails of some theoretical foundations, which are described in an introductory way in subsections~\ref{sec:parts}-\ref{sec:gran}. While they might appear `uninteresting' to some readers and some of it might seem irrelevant to computing, we include a few computing examples and we definitely will apply all that to computer programs in Sections~\ref{sec:vertical}-\ref{sec:whole} and therewith answer those three questions from Section~\ref{sec:intro}.\\

To meaningfully discuss the relevant details with respect to the scope of this document, some more preliminaries will be briefly summarised for the reader unfamiliar with them. They will assist in following the argumentation to establish that the computer program is a whole, and why. They are of a general educational or recap nature, as the case may be, rather than specifically IT only. In order to work toward a conclusion and justification, there are three relevant topics:
\begin{compactitem}[--]
	\item Part-whole relations, and in particular:
		\begin{itemize}
			\item[] 
			i) Mereological parthood, and possibly any of its refinements;
			\item[] 
			ii) Mandatory, immutable, and essential parthood; 
			\item[] 
			iii) Collective entities (collections) and sets.
		\end{itemize}
	\item Identity and unity.
	\item Granularity.
\end{compactitem}
After a basic note on computer programs and  the introduction of a running example, we introduce each one briefly informally for the reader unfamiliar with them, so as to keep the document self-contained. The respective formal characterisations (i.e., formalised in logic) can be found in the literature cited. For the most part, those details are not needed for the argumentation on software wholes, unless stated otherwise. 

\subsection{A note on computer programs and a running example}
\label{sec:programnote}

First, one could debate the definition of {\em computer program}. Consider, for instance, the definition of ``computer program'' as defined in the Copyright Act 98 of 1978 of South Africa\footnote{inserted due to 1 (g) of Act 125 of 1992 (available at 
\url{https://www.gov.za/sites/default/files/gcis_document/201409/act125of1992.pdf}; last accessed: 16 July 2020)}, 
it being: ``a set of instructions fixed or stored in any manner and which, when used directly or indirectly in a computer, directs its operation to bring about a result.''. This definition presumably entails that: {\em i)} it is essential that a computer program has some function that is `to bring about some result' x, {\em ii)} that it indeed can execute its task to bring about that result, i.e., the source code compiles (or is being interpreted) and successfully executes its task to perform function x, and {\em iii)} it distinguishes form from function, since it can be done `in any manner', i.e., {\em how} it achieves that function is beyond the definition's scope.  
The function view on computer programs is also held up in various courts cases \cite{Karjiker16} and, as we shall see further below, also from a theoretical viewpoint.

There are many types of computer programs. For instance, there is operating system (OS) software, such as Windows, Linux, and Mac OSX that have specific editions, such as Windows 10 and Mac OSX Yosemite 10.x, and utility software, such as disk defragmentation and activity monitor tools, which assist a user to manage the computer hardware. For end users, the most well-known category is applications, which are computer programs that require an OS to run to carry out its intended function. Popular examples of specific applications include Microsoft Word (a word processing program), WhatsApp for chat, and browsers, such as Google Chrome and Internet Explorer.  Among application software, there are further subtypes, such as desktop and mobile applications, and among those, further divisions, such as native mobile apps designed for a particular OS and hybrid mobile apps. Hybrid apps use Web technologies, such as HTML 5 and Javascript, and are then wrapped in a `native' wrapper specific for the OS of the mobile phone (Android, iOS etc.). A popular example of a native app is WhatsApp and example of hybrid apps are the Twitter, FaceBook messenger, and Uber apps. 

To answer the questions posed in the introduction, we need to dig a bit deeper, into the components that make up a computer program. Among the illustrations, we will reuse one multiple times, which is introduced in the following example

\begin{figure}[h]
	\centering		
	\includegraphics[width=0.8\textwidth]{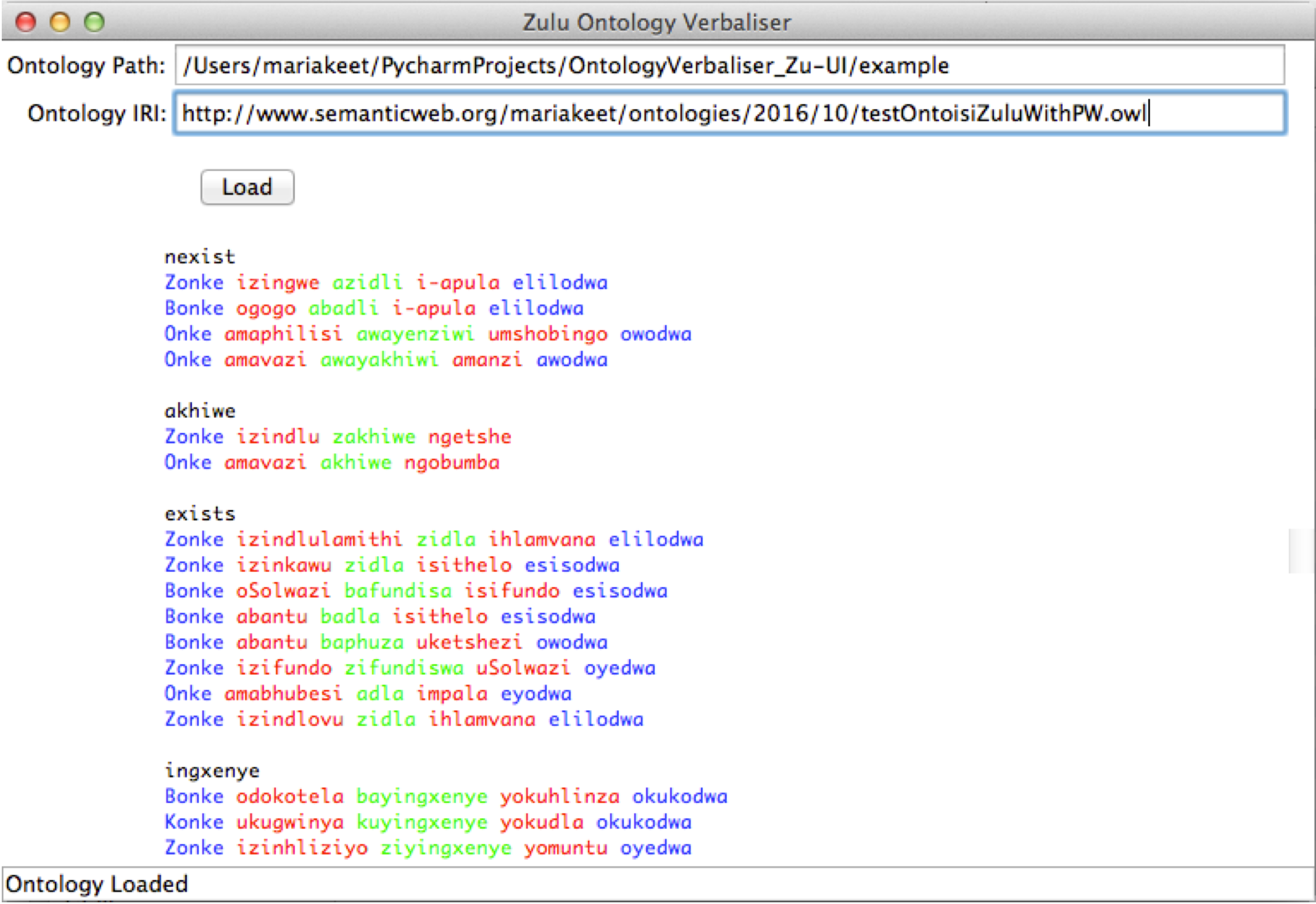}
	\vspace{-2mm}
		\caption{Screenshot of the isiZulu verbaliser with the enduser rendering of the output (source: adapted from \cite{KXK17}).} \label{fig:screenshotIVO}
\end{figure}

\begin{example}[Running Example: IVO]
As running example throughout this document, I will use the IsiZulu Verbaliser for Ontologies (IVO) \cite{KXK17}, which is a desktop application that puts into natural language (`verbalises') structured input in the form of an `ontology' that is stored in some text file (in OWL/XML format, to be precise) and that the author was involved in the design and programming of. Verbalising the OWL file into sentences in isiZulu is its only function. A snippet of the input in that OWL/XML format is as follows, which is normally {\em not} intended for human consumption, but for computer programs to process:
\begin{verbatim}
....
<SubClassOf>
        <Class IRI="#uSolwazi"/>
        <ObjectSomeValuesFrom>
            <ObjectProperty IRI="#fundisa"/>
            <Class IRI="#isifundo"/>
        </ObjectSomeValuesFrom>
    </SubClassOf>
...
\end{verbatim}
IVO takes such input and then generates natural language sentences from it in isiZulu (the Zulu language). The end user interface of IVO with the result of verbalising a particular ontology is shown in Figure~\ref{fig:screenshotIVO}, where the input file included the snippet above, which resulted in the third sentence in the {\tt exists} series of sentences, starting with ``Bonke oSolwazi...'', and the other sentences from the other serialised axioms in that file. It can also output that on the command line, but then without colours for the different categories of words. 
\end{example}

\subsection{Parts, wholes, and part-whole relations}
\label{sec:parts}

\begin{figure}[h]
	\centering		
	\includegraphics[width=0.95\textwidth]{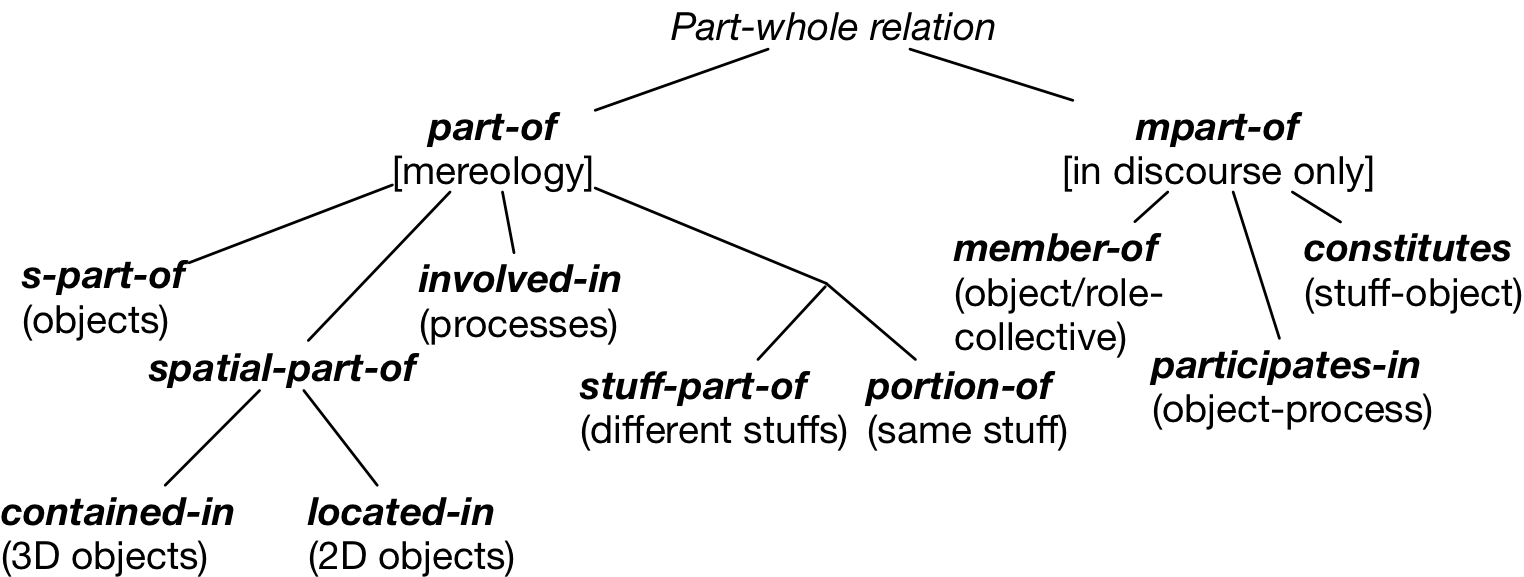}
		\caption{Summarised sketch of the part-whole taxonomy and informal descriptions of their domain and range (extended from \cite{AK07}/simplified from \cite{KK18fois}). 
		} \label{fig:potax6}
\end{figure}

Part-whole relations are used and are being investigated in several fields of study and applications, notably philosophy, ontologies for information systems, conceptual models for application development, linguistics, and natural language processing (see \cite{AK07,Varzi04,Winston87} and references therein). For instance, you heart is {\em part of} your body, in the process of eating food, there is a {\em part-process} of swallowing that food (and also taking a bite and chewing), and this paragraph is {\em part of} this section that is {\em part of} the document. 
There are some finer-grained details to it depending on the context where a `part of' is used in natural language, which has resulted in a fairly stable list of common types of part-whole relations, as shown informally in Figure~\ref{fig:potax6}, 
including mereological parthood, refinements thereof, and informal ones in natural language utterances only but not mereologically \cite{AK07}. 
For instance, instead of using aforementioned `part-process', one could name the parthood that relates only processes to its part-processes as, e.g., $involvement$ ($i$), specialising mereological parthood ($p$) that does not have any constraint on its participating entities; i.e., $\forall xy (i(x,y) \rightarrow p(x,y) \land process(x) \land process(y))$. Then, with a natural language utterance `each {\sf Eating} event {\sf{\em involves}} a {\sf Swallowing} event', one then identifies {\sf{\em involves}} as a surface realisation of the $involvement$ relation, and thus eating has as part swallowing. The easier way of communication is `eating involves swallowing'. 

\begin{example}[Part-whole relation examples for computer programs] Computer programs, when they run, can be seen ontologically as an ongoing process with sub (i.e., part) processes until someone closes the running of the program\footnote{Such a process in ontological sense is different from the technical meaning of the term `process' in computing, where it has a narrower meaning.}. Program code when the programmer writes it, is in letters such as these, and within that written `text', on can identify parts of the source code file, such as a `header', a class declaration, or a function, which can have components in turn. For instance, consider the function {\tt locpre} that adds a locative prefix to a word, such as adding {\em e-} to {\em iTheku}: 
\begin{verbatim}
def locpre(word,nc):
    phonolocword = 'e' + 'attached'
    if nc == '1a' or nc == '2a' or nc == '3a' or nc == '17':
        phonolocword = vowel_coal('ku',word)
    else:
        phonolocword = vowel_coal('e',word)
    return phonolocword
\end{verbatim}
It calls another function specified elsewhere in the code, being {\tt vowel\_coal}, that takes care of phonological conditioning when prefixing the word\footnote{Following the example: isiZulu does not have consecutive vowels, and the vowel coalescence in this case then applies to {\em e-} + {\em i-} of {\em iTheku}, which then returns {\em eTheku}. Locative suffix processing will eventually result in {\em eThekwini} (Durban).}. In turn, {\tt locpre} is called in the function {\tt pw\_ci}, which, in turn, is one of the functions to generate a natural language sentence (for {\em contained in}), which is a function in the IVO that has as function to verbalise an ontology. That is: one can construct a hierarchy (more precisely: a partonomy) of functions and part-functions. 

There is a notion of {\em atom}: those parts that cannot be divided further, i.e., that do not have any parts themselves. In the other direction, toward the ever greater whole, the boundary lies with the software specification: while our specification went as far as the IVO, one easily can take the source code and use it as a module in some grander computer program; e.g., one that accommodates the visually impaired and also synthesise speech of the sentences generated and that allows one to save both the sentences and the generated speech. What we see here conceptually, is some semblance of a software version of Lego, with basic building blocks, recognisable units, whole entities that have those units as part and that also can dual as modules (parts) of larger wholes, building software from components. 
\end{example}

Similar structural parthood relations can be observed in many other everyday artefacts. For instance: the cells of a heart are part of a heart are part of a body, the nail hammered in a piece of wood that is part of the bridge. These things are all physically connected. There are also whole artefacts whose components  are not physically connected, but stand in a part-whole relation nonetheless. For instance, a bikini and its parts, and if a manager orders a new ``desktop computer for the office'' for a new employee, it goes without saying that means machine + monitor + keyboard + mouse, not just the machine without any input/output devices. 

\paragraph{Collectives.}
On the right-hand side of Figure~\ref{fig:potax6} are part-whole relations that are not parthood in the strict sense of mereology.  For instance, in an sentence ``Ms. X is part of the project team GoTeam'', the `part of'  is better characterised as $membership$ or $grain$ relation between the part or grain {\sf Human} that Ms. X is and the collective {\sf Project Team} that GoTeam is an instance of, rather than parthood, 
 because transitivity cannot be guaranteed \cite{AK07}. A key feature of such constructions is that the object of the sentence is a physical object or a social object that is a {\em collective}, such as team, flock, or fleet. 
 For instance, {\sf Project Team} may be considered to be a collective consisting of human beings, any {\sf Flock} has as members sheep, and {\sf Fleet} has as members (grains) the ships that are part of the fleet. What exactly {\em collective} is, has been investigated as well, in fields as different as conceptual modelling for information system development \cite{Guizzardi05}, biomedicine \cite{Rector06}, and  social theory  \cite{Copp84}. They may endow collectives with features or constraints on the whole collective or on its parts. Widely-agreed upon key characteristics are that the collective: {\em i)}  has an  identity, {\em ii)} has some particular meaning, and  {\em iii)} it may survive its members, i.e., a member may change or be swapped but the whole keeps its identity. For instance, the rock band Queen can be identified, it has the characteristics of a rock band, and it has had some changes of membership along the way (it has had three bass players before John Deacon), yet Queen was already, and remained, Queen. 

One can put further constraints on the collective for it to be collective, such as that they can and do perform actions \cite{Copp84}, which is an argument popular especially in legal matters. Another possibility is to put an additional constraint on a collective's parts; e.g., that those parts must all perform the same role and if they do not, then the entity is more complex than a mere collective \cite{Guizzardi05}.  
Collective nouns in natural language, such as a fleet and flock thus may, or may not, be considered collectives ontologically, depending on such additional constraints.

\paragraph{Sets.} Collectives stand in contrast to mere sets. A set is just a mathematical `device' where a change in member changes the identity of the set, its members do not need a common binding feature, and therewith a set does not need to have a specific meaning, and sets have no agency. For instance,  your left foot and my laptop is a set with two objects as its members, but it is meaningless ontologically and it does not cause anything of itself. One may, of course, construct meaningful sets, such as the set of all mobile apps, but that does not mean that all sets have to be meaningful. Also, sets have members, not parts.

\paragraph{Particiaption in the relations.} 
In addition to looking at the type of part-whole relations, we can say something about how the part and the whole relate, which, perhaps, could be referred to as the {\em strength} of the participation of the part and the whole in the relation. 
There are four principal options \cite{AGK08}:
\begin{itemize}
\item[Optional:] the part $P$ may, or may not, be part of the whole $W$, or the $W$ may or may not have that part; e.g., a rear-view camera that may be part of a car, but a car does not need to have one.
\item[Mandatory:] some $P$ must be part of $W$, or vv., but not necessarily that one; e.g., for a house to be a house, it has to have a roof as part, but the roof can be renewed while still being the same house, and similarly for a human to mandatorily have a heart as part.
\item[Immutable:] that particular $P$ must be part of $W$ (or vv.), for as long as $W$ is in a particular role; e.g., a boxer must have as part his own hands and if he loses a hand he ceases to be a boxer (as entity he still may continue to live). 
\item[Essential:] that particular $P$ must be part of $W$ and if it loses that $P$, the $W$ will cease to be $W$; e.g., how the particular brain is part of a human: remove the brain and the human cease to exist as a human (turns into a corpse).
\end{itemize}
Besides how the part/whole participates in the relation, one also can make statements about shareability of the part:
\begin{itemize}
\item[Sharable concurrently:] some $P$ may be part of more than one $W$ at the same time, e.g., a talk may be part of a course and of a seminar series;
\item[Sharable sequentially:] $P$ is part of one $W$ at a time but can be part of different $W$s at different times, e.g., a removable network card may be part of one PC at one time and part of another one another time but it cannot reside in two different PCs at the same time; 
\item[Exclusivity:] $P$ can be part of at most one $W$, e.g., entities such as the brain and spinal cord as part of a human body 
and some particular swallowing event that was part of eating your lunch yesterday.
\end{itemize}

\paragraph{Functional parts.}
To talk of functional parts, we first have to consider {\em function}. Rather than investigating that here, we use an off-the-shelf definition that emerged from extensive research. Mizoguchi {\em et al.} define functions as ``a role played by a behavior specified in a context'' (for detail, see \cite{Mizoguchi16}). While this definition may sound somewhat obscure, it is closer to the meaning than, say, the main Oxford Dictionary definition\footnote{It defines function as ``An activity that is natural to or the purpose of a person or thing.'', but a function need never be realised so then is not an activity that is happening; e.g., after using the new scissors to screw in the screw, it may never perform its intended function of cutting because the blade got bended. (Definition's source: \url{https://www.lexico.com/definition/function}; last accessed on 14 July 2020)}, since this is broad enough to also include situations where, say, a pair of scissors is used to screw in a screw, in that that pair of scissors performs the role of screwdriver. Many types of function have been identified; e.g.,  the ontology of function identified 89 different types of functions \cite{Mizoguchi08}. Two key observations are that a ``function is necessarily supported by the structure and/or properties of the things'' and that ``one of the most significant properties which function must have is implementation-independence.'' \cite{Mizoguchi08}. 
This thus relates `vertically' between the part and the whole, where those parts contribute to realise the overall function (with to without component functions that contribute to that) and the structures that realise that  would thus be ``functional parts'' of the whole in some way \cite{Mizoguchi17,Vieu05}. 
The definition of {\em functional part} provided by Mizoguchi and Borgo \cite{Mizoguchi17} is as follows:
``Given an entity A and a behavior B of it, a functional part
for that behavior is a mereological part of A that, when installed in A, has a behavior that
contributes to the behavior B of A.''
This is underspecified when taken in isolation, because it is meant in the context of their definition of function of the ``a role played by a behavior specified in a context''. Therefore, we refine the definition as follows:
\begin{definition}[Functional part]\label{def:funcpart}
Given an entity $x$ and a behaviour of type $B$ of it that plays the role of function F in that context $c$, a functional part for that behaviour of type $B$ is a mereological part of $x$ that, when installed in $x$, has a behaviour of type $D$ that contributes to $B$ of $x$, where $D$ plays the role of part function $F'$.
\end{definition}
Mizoguchi and Borgo then go on to identify four types of functional parts which are: genuine, replaceable, persistent, and constituent functional part. 
A part is a {\em genuine} functional part if  it is installed correctly in $x$, from a structural viewpoint; a {\em replaceable} functional part refers to the regular `mandatory' participation, discussed earlier; a {\em persistent} functional part is either an essential or an immutable part; and a {\em constituent} functional part is a generic part regardless its assigned position as it may be temporarily taken out physically \cite{Mizoguchi17}. 
Although Mizoguchi and Borgo did not formally characterise these parts, it is clear from the context in the paper that it is to be understood as {\em proper} part. 

\subsection{Identity and unity}  
\label{sec:unity}

Ascertaining the identity of an entity is about examining something as being the same or different at one point in time (synchronic identity) or across time (diachronic identity). To establish the identity of an entity, one uses {\em identity criteria} that the entity has, which is a stronger notion than {\em identification criteria} (typically a single artificial attribute) that  may be assigned to entities. For instance, one may identify {\sf\em Tibbles}, the neighbour's cat, by the colour of its fur, its size, and how it meows, whereas artificial attributes of entities assists with identification, such as one's ID number and the MAC address of a network card. Within the context of software, it may be of use to also note ``striking resemblance''  or ``comprehensive non-literal similarity'' of program code\footnote{The famous  cases where this got established are Franklin vs. Apple (1984) where Franklin had copied Apple's operating system often line by line 
and Whelan vs. Jaslow (1987) where Whelan had developed a program in Fortran and Jaslow had then rewritten it line by line in Pascal.}, 
which holds when two pieces of code are identifiably different, but {\em the semantics/logic of the expression of the idea} is the same. Two typical examples are 1) to take some piece of code and change the names of the variables: they are different documents with, to the letter, different text, but have striking resemblance; 2) to keep all the business logic of the code but translate it line by line from one programming language into another language (say, Java to C++). 

Unity \cite{Guarino00a,Guarino09oc} is a special case contributing to identity, which focuses on the relation that the parts have that make up the whole entity that has some identity.  {\em Unity criteria} are those criteria that have to hold among the parts for the whole to be a whole. Depending on the nature of the unifying relation, one can identify different types of wholes. Some typical examples are {\em topological wholes}, such as a tree and a heap of sand, {\em morphological wholes}, such as a bouquet of flowers and a constellation of stars, {\em functional wholes}, such as scissors and a bikini, and {\em social wholes}, which are certain types of collectives, such as a population of a country and a herd of sheep. Besides the aforementioned part-whole relation between the parts and the whole, there is a `something' that binds the parts together to be a meaningful whole with an identity, which stands in contrast to, notably, stuffs (typically indicated with mass nouns, such as water, mayonnaise, 
and wood) and arbitrary sets of entities like the set of your left foot and my laptop. Since in the current context 
we only deal with programs, we exclude stuffs or amount of matter from the possible relata henceforth. 

Guarino and Welty sought to formalise the notion of unity, with a good start made in  \cite{Guarino00a}: 
with $B$ the generic unifying relation that binds the parts to each other with respect to the whole, and $P$ mereological parthood that is temporally indexed (presumably with time $t \in \mathcal{T}$ as discrete time), their first condition is that
\begin{equation}
\forall xyzt (P(x, y,t) \rightarrow (P(z,y,t) \leftrightarrow B(x,z,t))) \label{eq:unity}
\end{equation}
Informally: if there is a parthood between $x$ and $y$ as some time $t$, then at the same time it holds that for any entity $z$ that is also part of the same whole at that time, then it is in a unifying relation with the part $x$, and vice versa.

Eq.~\ref{eq:unity} and the description in \cite{Guarino00a} assume weak supplementation would apply, under the assumption that $x \neq z$, which they specified as $\forall xy (PP(x,y) \rightarrow \exists z (PP(z,y) \land \neg O(z,x)))$, i.e., every proper part $PP$ must be supplemented by another, disjoint [not overlapping $\neg O$], part, 
because of the implication: there must be another part of $y$, being $z$, so that they can relate through $B$. 

One would want to exclude any mereological sum that is just a contingent whole, like my thumb and index finger. This can be seen in analogy to the difference between mere sets and collectives. Noting a typo in their formalisation in \cite{Guarino00a} (corrected in their lecture slides on OntoClean), and adding quantification, we formalise it as 
\begin{equation}
\neg \forall xzt (B(x,z,t) \leftrightarrow \exists y (P(x,y,t) \land P(z,y,t))) \label{eq:notarbitrary}
\end{equation}
That is, it does not hold for all $x$ and $z$ that when they are related through $B$ at time $t$ they are also a part of $y$, or: $x$ and $z$ may be bound in some way, but they are not therefore also part of a whole $y$ because of that. This raises a few questions about the exact configuration when Eqs.~\ref{eq:unity} and \ref{eq:notarbitrary} are taken together. For instance, can or do they have to be different  $B$s that relate $x$ and $z$? Can they be part of different wholes, or: are the $y$s in Eqs.~\ref{eq:unity} and \ref{eq:notarbitrary} supposed to refer to the same whole (which they currently do not)? Weak supplementation aims to rule out the possibility that $x = z$, in that there cannot be a whole with a single {\em proper} part, but unity is declared with $P$, not $PP$, so would unity still apply just in case there is only one part? These finer-grained details about the configuration brings us to the last preliminary: granularity.

\subsection{Granularity} 
\label{sec:gran}

Granularity refers to the level at which one operates or analyses and that there are different levels of granularity \cite{Keet08phd}. For instance, at level $L_i$, some {\sf amount of water} is a type of stuff, but at its finer-grained level $L_{i-1}$, there is a set or collection of {\sf H$_2$O molecule}s, or at $L_i$ one has an instance of {\sf Team} and at $L_{i-1}$ are its grains, {\sf Human Being}s. The level of analysis may be of importance ontologically. For instance, {\sf Organisation} is a social entity at level $L_{i}$ that has one or more instances of {\sf Human} (more precisely: its employees) as members at $L_{i-1}$. It is only the former, however, that must have a bearer with a single physical address that holds for that entity at that level. That feature is not inherited by its constituent parts, since, e.g.,  the company's employees typically live elsewhere. Similarly, it is the app that has, at that level of granularity $L_i$, the property of directing the computer's operation to bring about a result. That is, the whole entity at $L_{i}$ has properties of its own, which can be---and typically is---very different from what its constituent parts do at the finer-grained level $L_{i-1}$ where they reside. 

Note that granularity focuses on the more/less details of analysis, which is different from prioritisation at the same level of granularity. For instance, there may be direct parts $x,y$, and $z$ at level $L_{i-1}$ of whole $w$ at $L_{i}$ and one is interested only in $x$ and $y$. This does not push $z$ into a finer-grained level $L_{i-2}$, since an ontologist's 
interest is a separate matter.

\section{Software components and their relation to the source code and program}
\label{sec:vertical}

Let us first have a look at some important features of computer programs, which concern their parts, construction and use, and how those files relate to source code and executable. 
\begin{figure}[b]
	\centering		
	\includegraphics[width=1.0\textwidth]{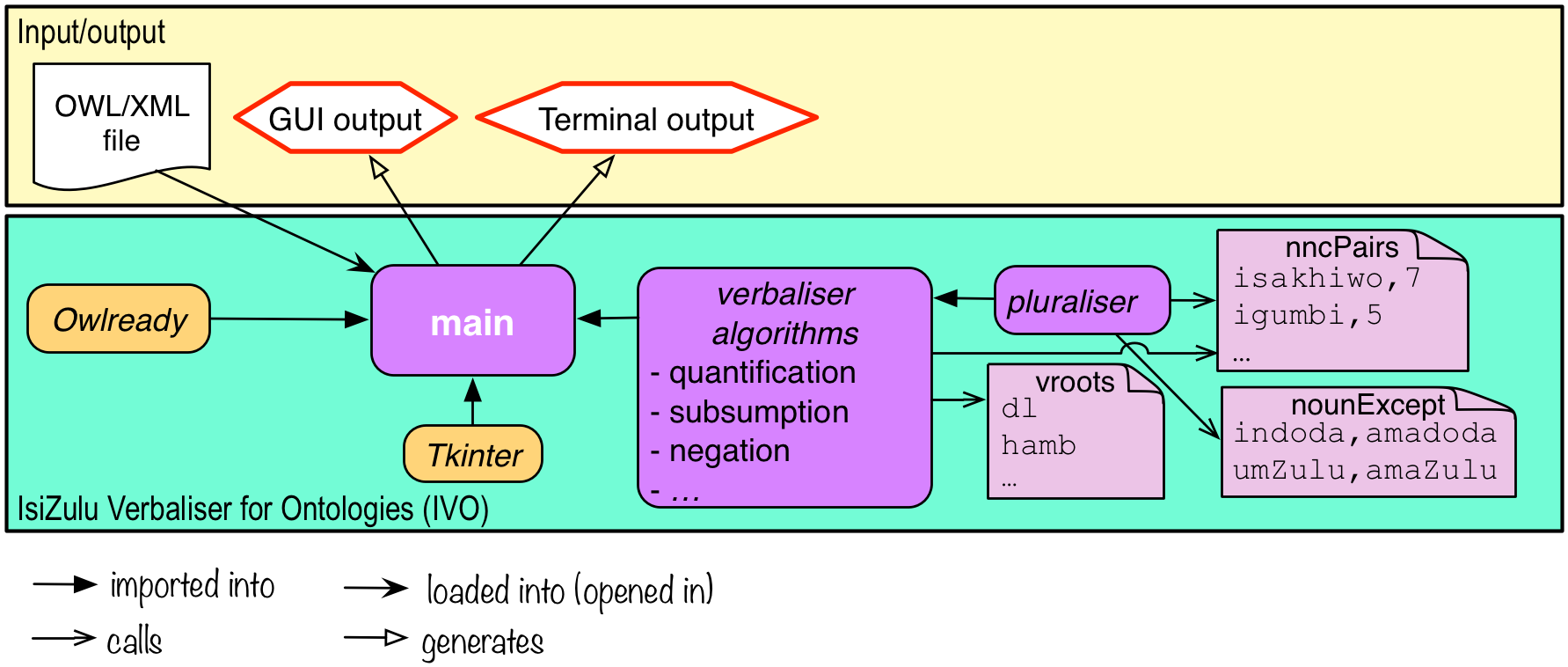}
	\vspace{-2mm}
		\caption{Principal components of the OWL verbaliser (source: adapted from \cite{KXK17}).} \label{fig:arch}
\end{figure}
For illustrations, we continue to use the IVO, but now we start looking `under the hood' to its architecture and components, whose principal components are shown in Figure~\ref{fig:arch} and described  in the following example.

\begin{example}[IVO's architecture]
Consider Figure~\ref{fig:arch}, with the coloured shapes: the large green rectangle is the IVO computer program, with {\sf main} being the central component that is used to `launch' it and to do some final processing. {\sf main} pulls all the other components together, being 1) the integrated modules (in orange) {\sf Owlready} and {\sf Tkinter} that already existed beforehand and were developed by someone else, 2) the two other clear purple boxes with program code that this author wrote ({\sf verbaliser algorithms} and {\sf pluraliser} of nouns), which, in turn, 3) call certain text files (pale purple, right-hand side) with required data structures to function properly. The yellow rectangle separates out the input/output: one gives IVO a text file in OWL/XML format, to load and process, IVO will then process it and return the generated sentences as output either in the GUI, like in Figure~\ref{fig:screenshotIVO}, or in the terminal. 
\end{example}

One has to distinguish between the {\em source code} as artefact and the {\em compiled code} (or interpreted code) as artefact, since multiple source code files may compile into one compiled artefact that is a digital file (the executable), 
$x$ source code files may result in $y$ (with $x \neq y$) files of the software application, and the application has one `point of entry' to launch it irrespective of the number of files it needs for its proper operation to bring about a result. Since the arguments are the same for compiled code and interpreted code, we will use for convenience just the shorthand of {\em compiled code} and {\em compilation process} henceforth, but it all just as well applies to program code written in interpreted languages, such as Python, and who's program files are interpreted into machine code as well.

\subsection{Modularisation and compositionality}

Good design principles, be they for engineering for physical objects or for software, avail of {\em modularisation}  and {\em compositionality} \cite{Bjorner10,Noback18,Prehofer08,Tripakis16}. That is, one breaks up a large problem into smaller ones, devises solutions for the smaller problems, and  puts those together---declaring relations 
or rules-as-relations to 
specify how the components interact---to solve the larger problem. Some key benefits of this approach include manageability, division of labour when multiple people work on solving the problem, potential for reuse, quality control, and maintainability. 

\begin{example}
For instance, with IVO (Figure~\ref{fig:arch}), {\sf Owlready} is reused from elsewhere, two files with the algorithms access the same data structure file ({\tt nncPairs.txt}) rather than having to declare that twice (once in each file), and the vocabulary is easy to maintain by just updating a text file rather than having to dig through the algorithms in the source code files with program code.
\end{example}

There is a way how these files relate to the program and source code as a whole, and, taking a small abstraction step from concrete files to conceptual containers, how components of the architecture of the software relate. Since they are components, one should have a look at theories of how components, i.e., parts, relate a whole: this is the topic of {\em mereology}. A mereology-based approach (joined with category theory to accommodate multiple types of part-whole relations), 
has been applied to component-based software architectures \cite{Le08}, but not at the level of code. One could extend that idea by assuming that each component of the architecture is realised by one file. Practically, especially for production-level software, each component of the architecture may be divided up into smaller components to realise the purpose of the component, and some may be combined into one file, which brings us to the next point.

There is a difference between {\em what} an artefact, such as a computer program, is designed to do, i.e.,  its function or intended purpose, and {\em how} that is realised. This follows directly from the modularisation, since there may be multiple ways of breaking up the design into smaller components. Indeed, for some simple artefact, such as a hammer, these options are rather limited. A straight-forward webpage can have everything declared in one single HTML file, but that also can be split up into multiple files: the text to be displayed written in an {\tt .html} file, the layout relegated to a style sheet in {\tt .css} called from the {\tt .html} file, and some javascript in a separate {\tt .js} file also called from the {\tt .html} file. This can be much more complicated for programs, and especially in their source code form. 
There may be modularisation by sub-function that each are stored in a separate file.

\begin{example}
Regarding sub-functions, consider the {\tt pluraliser.py} that generates a plural from a noun in the singular that is used by the module that generates the sentences, imported packages that themselves consist of several files (e.g., {\sf Owlready}), a data structure that is essential to several functions that are written in several files and therefore it is stores in a separate file (e.g., {\tt nncPairs.txt}), a file with configurations values, and so on. 
\end{example}

Then, while a single source file in, say, the programming language C will result in a single program file, in the intermediate steps there are other files and processes involved: 
during the compilation process that translates the source file into the object file, {\tt .lib} files are called by the linker to generate the program file. 

In this context of function and modularisation, let us digress briefly into the {\em user experience} of the computer program. A  function may be realised by more or less well integrated components that may be automatically invoked or manually started. Manual actions may give the user the impression of it being distinct but related components, whereas automated processes will provide the impression of it being one entity. This may be configured as such by design. For instance, a user may by default block the running of scripts in spreadsheets, so when there is one, the user has to approve of its execution. However, when the security setting is such that running scripts is permitted by default, end-users will not be aware that it is executing the script as a sub-process of the spreadsheet process or even that there is a script running at all. 

User experience of seamless integration is not a good measure whether some component file is really a part of the program, or whether there are actually multiple programs. A possible way of determining that, is with respect to the code and behaviour\footnote{e.g., in the GNU FAQ about ``aggregates'': \url{https://www.gnu.org/licenses/gpl-faq.en.html#MereAggregation}; last accessed on 16 July 2020. \label{fn:gnu}}: how do the files interact (i.e., what mechanism), and what does that interaction mean (i.e., the semantics of the communication)? The GNU FAQ suggests that links such as function calls and inclusion by reference definitely is a tight enough interaction for indicating parthood, as are the links resulting from the linking stage in the compilation, whereas piping on the command line---e.g., output from one to be used as input into another---is seen as too decoupled interaction and therewith a reason to lean toward them being separate programs, with the caveat of ``exchanging complex internal data structures'' where the semantics of what is being communicated is ``intimate enough'', then it still may be considered as one program with two parts. Either way, these are examples of considerations for possibly counting as unifying relation that is required for a whole to be a whole (recall Section~\ref{sec:unity}). We shall return to this, and attempt to resolve that aspect of interaction among (possible) parts, in Section~\ref{sec:horizontal}. Currently, to the best of our knowledge, there are no formal, widely agreed-upon, and standardised criteria for determining whether a file is part of a program or not.

\subsection{Mode of participation in the part-whole relation}
\label{sec:participation}

We have to consider in what way the files of a program participate in the relation to the program as a whole. As noted at the end of Section~\ref{sec:parts}, there are four options: optional, mandatory, immutable, and essential participation. We now have to assess which one(s) apply in which cases with respect to the source code and the compiled program. On principled grounds, we will first take this to be under the assumption that the program is free of bugs, and sources code that needs to be compiled, will do so without errors. 

{\em Optional participation of a file:} In the interest of optimal design and development and optimised (minimised) software, there should be {\em no} optional files either in the source code or in the running program. That is: if the source code still compiles and the program can continue to operate without any bugs without that file, then that file is redundant and should be removed. Yet, there may well be sub-optimal source code and programs, in particular when some modules or libraries are imported. Such imports are typically used as-is and operationally redundant files may not have been removed to save oneself testing whether it could be removed.  

\begin{example} IVO and optional participation: 
The programmers did not check whether all the files of the imported {\sf Owlready} are used and just bundled {\sf Owlready} with the verbaliser. It is very likely that not all {\sf Owlready} source files are used, since IVO only fetches data from the OWL file but does not modify it, yet there are functions stored in separate files to modify the ontology, such as {\tt instance\_editor.py}, which is never used by IVO at least. 
\end{example}

{\em Mandatory participation of a file}: the file has to be present and at the right location, but it can be swapped for another without altering the function and functioning of the software and it won't introduce bugs. This would be the most common case.

\begin{example} Mandatory participation applies to IVO as well: the code looks up nouns in the {\tt nncPairs.txt} file for any axiom it has to verbalise. If it is not there, it will throw an error, but it can be overwritten with a different file also called {\tt nncPairs.txt} that has to same outward function and minimal (but possibly more) content, e.g., one that has more nouns with noun classes listed in the file. The function---verbalising the supported types of axioms in isiZulu---remains the same regardless. 
\end{example}

{\em Essential participation}, in short: remove that particular file and the app is `broken'. There are different levels of `broken', however, such as having minor bugs but the core function still can be achieved mostly or fully, major issues so that it only works partially, and the program not being able to run at all. From the binary viewpoint of `bug-free or not', these differences do not matter: if not all specified functions, including sub-functions, work when {\em that particular file} is not present, then it is an essential file. 

\begin{example} Essential participation may hold for the {\sf Owlready} module. For instance, if any other version than that instance {\tt Owlready-0.3} is not backward or forward compatible, then that version is essential to the function of verbalising ontologies by IVO.
\end{example}

{\em Immutable participation of a file:} putting the definition in terms of software, then when the software performs a particular role, it is essential to it. It is highly 
debatable whether this mode of participation in a part-whole 
relation 
would be applicable for either source code or computer program, for two main reasons. First, if at some point the program has to perform some function $x$, then it is part of the specs at all times. The only case where it may be argued to be applicable is when the program has multiple alternate functions, and one is using only a subset of it, then if there are files essential to the operation of that subset of functions but they are not used when running deploying the tool differently in a different, disjoint, function, then they could be said to be immutable parts. The easy counter-argument to that, however, is that the software specs has the requirement that the software has to have those two different functions, and then we're back at mandatory participation at least. Second, in addition to the former, for that particular file in that particular role or mode of operation, it would have to be the case that only that version of the file or module would work with the rest of the program.

\begin{example} IVO has two `modes' of operation---terminal-based with as aim to incorporate the verbaliser into other tools and one for end user interaction for reading of the output 
with a GUI. {\sf Tkinter} is only essential for the GUI operation mode\footnote{Tkinter is the {\em de facto} user interface module for Python; see also \url{https://wiki.python.org/moin/TkInter}; last accessed on 16 July 2020.}, but not the terminal-based mode that still performs the essential function of verbalising ontologies. One thus might argue that {\sf Tkinter} could be an example of an immutable part of IVO, if it were the case that another version of {\sf Tkinter} would not be able to generate the user interface. Yet, if the requirement is `to be able to operate in two modes, terminal and  with end user formatting', then omitting {\sf Tkinter} breaks the overall functionality, and would therewith be at least mandatory (if not essential), rather than immutable. The requirement was to have both ways---the GUI for the linguist to check the output easily and the terminal output in anticipation of reuse---so then immutability does not apply in this case.\\
\end{example}

In sum, for bug-free operation, mandatory participation in the parthood relation 
is the common case, with optional participation arguably amounting to lazy or time-constrained coding and essential participation would be an example of `brittle' coding practices. With these principles in place, one can start going off a sliding scale: what if we let go of the bug-free premise? What if one prioritises the functions of a program? Drawing the line in that case will end up to be subjective and dependent on the example. For instance, take the relatively well-known program MS Word: even if the Clip Art feature does not work, it still easily can go on with word processing, but if the `insert table' or `create columns' functions are dysfunctional, definitely less so, although one still can type and save text. This would require  establishing a partonomy of functions, each with priorities or weights assigned, and some threshold function to compute the outcome. \\

Lastly, {\em shareability of a file} depends on whether a file has to have a lock on it when open because of concurrency issues. If that is the case, then that component file of the computer program (not the source code) is only sequentially sharable. Due to the unclear nature of software artefacts, exclusivity is debatable depending on the choices---among others,  whether copies count as distinct and whether it is the file as information artefact, etc. Since this is not crucial for source code copyright/copyleft issues, trade secret violations, and illegal downloads, we will not address this topic further.

\subsection{Functional parthood}
\label{sec:fp}

Since there is the compositionality thanks to modular design, and the types of files with their contents are distinct rather than subsets, the applicable relation between those component files and the whole---the {\sf Source code} or its compiled (or interpreted) {\sf Executable}---is that of {\em parthood}. One may argue that it would also be a {\em proper} parthood in most cases, because production-level software normally consist of multiple files in source code and the compilation process. 
Multiple part-whole relations have been proposed (recall Figure~\ref{fig:potax6}), so this relation possibly may be refined. However, because the ontological status of a file and of a computer program is not fully settled \cite{Turner19}, a confidence in applicability of a refinement along the line of \cite{AK07} is limited, except that from a `written text viewpoint' on code, it would probably be {\em structural parthood} and when the program is being executed there are running processes and sub-processes, hence, {\em involvement}. 

It is complicated further in that a program has a specific assigned function, such as {\sf Text processing} for {\sf\em MS Word} and {\sf Song management} for {\sf\em iTunes}, which each have sub-functions, such as {\sf Text formatting}, {\sf Printing}, and {\sf Playlist creation}, i.e., there are  functions and part-functions, and they in turn may have part-functions, such as {\sf Changing text to bold face}, {\sf Adding song to playlist} etc. 

Recalling Definition~\ref{def:funcpart} for functional parthood, and the four types of functional parthood that Mizoguchi and Borgo specified \cite{Mizoguchi17}, we can no apply this to clarify the relation between the parts of the computer program and the program. If we assume that source and compiled code consist of more than one file and there are no optional files, then 
\begin{compactitem}
\item Source code of the computer program: all files are genuine functional parts and replaceable. 
The files have to be in specific locations to be called.
\item Compiled code of the computer program: all files are genuine functional parts and possibly also persistent (if not, then they are replaceable). 
The files have to be in the right folders and only there. 
\end{compactitem}
This finalises the `vertical' relation between the parts and the whole. We will move on to the relation among the parts in the next section.

\section{The relation that binds the components}
\label{sec:horizontal}

Taking stock, the files that are part of a computer program participate mostly mandatorily in the genuine functional parthood relation ($genuineFP$) with the source code or compiled program, normally there is more than one file involved, and each file either contributes to some function or performs one or more functions to contribute to the specified function(s) of the software. 
In terms of applying Eq.~\ref{eq:unity} toward sorting out that `unity' for wholes, a first step would thus be the following:
\begin{equation}
\forall xyzt (genuineFP(x, y,t) \rightarrow (genuineFP(z,y,t) \leftrightarrow B(x,z,t)))  \label{ex:gFP}
\end{equation}
Note that the time indexing with $t$ is kept here from Guarino and Welty's Eq.~\ref{eq:unity}. This is more precise as it allows for specification whether it has to hold at some time or at all times or at different times, rather than not saying anything at all about it. For computer programs, it may assist later on with any possible extensions that would take into account versioning, without affecting the foundations.

This brings us now finally to addressing the research questions posed in Section~\ref{sec:intro}. First, we look at its question Q2, which amounts to resolving the ``unifying relation'' $B$ for the source code and the compiled program, and related loose ends from the formalisation of unity in Section~\ref{sec:prelim}, such as the implicit assumption that $x \neq z$ in the axiom.  

\subsection{The unifying relation $B$: can it even be applied?}

When we characterise $B$ from Eq.~\ref{ex:gFP}, i.e., the relation among the parts, it will ensure we arrive at either that the program is a (functional) whole since it has unity, or, if it cannot be specified, that then there is no whole after all. 
What such as $B$ then may be  is unclear from the expos\'e on unity in \cite{Guarino00a}. There is not even one example of such a  unifying relation and nor do the OntoClean examples 
\cite{Guarino09oc} have any example for $B$ either---it is just assumed to exist with some appeal to intuition.  
Philosophers, such as \cite{Copp84,Koslicki13}, do not do a better job at it, like with a mere ``x is unified in the right way'' \cite[p180]{Koslicki13}. At the other end of the spectrum in the conceptual modelling area, it goes the other extreme with non-reusable situation-specific instance-level relations like ``carrying out research in the same sub-area of the area of distributed systems in the University of Twente'' \cite[p159]{Guizzardi05}, but no type-level relation that may hold throughout similar cases. 
For instance, while {\sf Bikini} is a functional whole consisting of two parts, the particular type of unifying  relation that holds between the two parts, for any bikini, is left undetermined.   
However, if we are to assess program, then this unifying relation needs to be ascertained, for it would assist explaining {\em why} it is a whole or not. We shall try this in the remainder of this section.

Let us first consider two cases with commonsense examples. For a morphological whole, such as the {\sf Constellation} example (e.g., the stars making up the {\sl Sagittarius} sign), this may be, say, that all the stars involved share that they participate in the {\sf connecting the drawing lines} to make up the figure that is a certain shape. For social wholes, such as {\sf Electorate} of a country, its parts (or members of the collection)---the voters---share that that they stand in a {\sf fellow adult citizen} (of some country) property relation to each other. Applying Eq.~\ref{eq:unity} specifically for the purpose of illustrating its workings, let's formalise the latter as follows:
\begin{equation}
\forall xyzt (VoterIn(x, y,t) \rightarrow (VoterIn(z,y,t) \leftrightarrow fellowAdultCitizen(x,z,t))) 
\end{equation}
So, if The Daily Show presenter {\sf\em Trevor Noah}  is a voter in the {\sf\em Electorate of South Africa}, then if there is another voter, say, the national rugby team's captain {\sf\em Siya Kolisi}, in the {\sf\em Electorate of South Africa}, then {\sf\em Trevor Noah} and {\sf\em Siya Kolisi} are fellow adult citizens and vv.; hence, it can be applied generically and for specific instances.

\subsection{$B$ for computer programs}

Since the notion of unity seems to be usable at least in other domains, let's consider programs: 
what is it that unifies, or binds, the files, other than some vague `being another component'? Because there is a compiled program and a source code program, with as consequence that the files comprising each are different, the relation among the file is, perhaps, different, too. Therefore, we shall address each in turn. In both cases, we use the `realistic case' argument, i.e., setting aside the corner case of one main file without calling anything else ever, since we are interested in practical applications.  

\subsubsection{The unifying relation for source code files}

Recall that regarding source code, 
the whole is the source code for the application, such as {\sf\em Firefox v77.0 source code} or {\sf\em Linux kernel v5.7 source code}, which operates at $L_i$ level of granularity, regardless how that is organised. The components and the organisation thereof---i.e., the files that make up the source code like a file with the {\tt main()} and imports---operate at the $L_{i-1}$ finer-grained level of granularity. This is likewise for the computer program, such as the {\sf\em Firefox executable app} and {\sf\em Linux kernel} with the compiled code.

\begin{figure}[t]
	\centering		
	\includegraphics[width=0.9\textwidth]{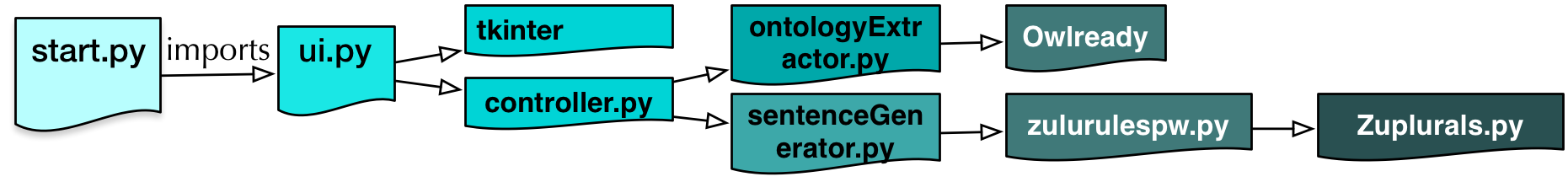}
	\vspace{-2mm}
		\caption{Series of actual {\tt import} statements of the isiZulu OWL verbaliser, for the Python files only (it calls other files as well, not shown).} \label{fig:import}
\end{figure}

There are various ways to link the source files to the `main' file that stands at the root. For instance, there may be {\tt import} statements, as is the case with the Python and Java programming languages, an {\tt \#include} statement in case of C and C++, or some other coupling mechanism, such as function calls \cite[ch9]{Noback18}. 
An example of cascading imports for the IVO source code is depicted in Figure~\ref{fig:import}. The organisation of the source files form a tree that may have files with more than one parent, which is also called a dependency graph. Since each coupling mechanism goes one way, it is called `directed'. Mathematically, it is thus a directed graph with a top-node, where the other nodes may have more than one incoming edge and more than one outgoing edge, ideally that graph is acyclic, and through that chain of connections one should be able to end up at that top. A circular import, be it directly between two files importing each other or indirectly through a chain of import statements 
may not be forbidden by the syntax of a programming language, but it is considered an anti-pattern, i.e., bad design, and, consequently, there is an {\em Acyclic Dependencies Principle} with strategies for breaking cycles to foster good design \cite[ch9]{Noback18}. 
Also, there may be `orphans' in the source code: one or more files that do not relate to anything, or at least do not seem to. While one cannot guarantee they are useless during execution, it is highly likely to be either excess baggage or the coupling mechanism is so loose, it may have to be considered to be a separate program. This argument is advanced also by GNU's copyleft licences (cf. footnote~\ref{fn:gnu}). 

Assuming the aforementioned sufficiently tight coupling where the files will show up in a dependency graph, then, for 
any two component files of the source code, being genuine functional parts, they thus relate through that directed graph of dependencies, either further down in the hierarchy or upward and into another branch. One thus always can construct a path between any two nodes (ignoring the direction of the edge). Let us call that the $SCgraphPath$ property, then Eq.~\ref{ex:gFP} can be updated into Eq.~\ref{ex:gFPTP}:
\begin{equation}
\forall xyzt (genuineFP(x, y,t) \rightarrow (genuineFP(z,y,t) \leftrightarrow SCgraphPath(x,z,t)))  \label{ex:gFPTP}
\end{equation}
That is, there is a notion of unity for source code. 
We still need to verify it is not just a contingent whole. That is, either the following holds:
\begin{equation}
\neg \forall xzt (SCgraphPath(x,z,t) \leftrightarrow \exists y (genuineFP(x,y,t) \land genuineFP(z,y,t))) \label{eq:notarbitraryFPTP}
\end{equation}
or, at least, that it holds with Eq.~\ref{eq:notarbitrary}'s generic $P$ rather than the refined $genuineFP$. 
Is it possible that $x$ and $z$ are (genuine functional) parts of some whole $y$ but not do stand in a  $SCgraphPath$ relation to each other, or the other way around? This is indeed possible:  they may be parts of a library $y$ of header files that come with the C/C++ Integrated Development Environments  such as Eclipse\footnote{\url{http://www.eclipse.org/cdt/}; last accessed on 16 July 2020.} or $x$ and $z$ are in a Python repository of modules $y$, such as PyPI\footnote{\url{https://pypi.org/}; last accessed on 16 July 2020.} or your own one. In those cases, $x$ and $z$ are neither in the $SCgraphPath$ relation nor, arguably, $genuineFP$ but another part-whole relation, since there they are part of a collective. It may be that one header file imports another, or one module avails of another in such as repository or library, but this need not be the case, and many files will not do so.

Lastly, Eq.~\ref{eq:unity} implicitly assumed weak supplementation (`there must be some other part, too')  even though it was not formalised as such. That is, when applied to the software case, that there ought to be more than one source code file. There exist source code that has no {\tt import} or {\tt \#include} statement, although this is rare for production-level software. Still, with the current formalisation, even such a corner case is not a problem, since 1) weak supplementation is asserted for $PP$, not $P$, but $P$ is used in the unity axiom and 2) for ``$\neg \forall xz$'' in the axiom to hold, even just one example satisfies it, which we just have with the module library. 

We therefore can define source code to be a functional whole, as follows:

\begin{definition}[Program source code as a functional whole]\label{def:scdef}
For any computer program $C$ in source code form that has one or more genuine functional parts $x_1, \ldots, x_n$ with $n \geq 1$ and $1 \leq i,j \leq n$, and 
time $t \in \mathcal{T}$ where $\mathcal{T}$ is the set of discrete time point, it holds that 
$\forall x_ix_jyt (genuineFP(x_i, y,t) \rightarrow (genuineFP(x_j,y,t) \leftrightarrow SCgraphPath(x_i,x_j,t))) $.  For at least some $y$, $SCgraphPath(x_i,x_j,t)$ does not hold, i.e.,: \\
$\neg \forall x_ix_jt (SCgraphPath(x_i,x_j,t) \leftrightarrow \exists y (genuineFP(x_i,y,t) \land genuineFP(x_j,y,t)))$. 
\end{definition}

This concludes the part of the argument that, besides the `vertical' relation of (functional) parthood, there's also that `horizontal' relation among the parts with the dependency graph counting for unifying relation for the source code.

\subsubsection{The unifying relation for the object code (compiled program)}

How can Eq.~\ref{ex:gFP} be finalised for the computer program, as in the executable (object code) version of it? At that stage of the software development, the explicitly stated to be imported files have been included, so $SCgraphPath$ is not applicable here and thus no tree of source code files, since they have been compiled.  
Yet, there still may be multiple files that are part of the  computer program that ``brings about a result''; examples include an {\tt .ini} file for initialisation setting, an {\tt .xml} file with the locale setting for menu options in a different language, a dictionary to load for spellchecking, and a {\tt .dll} as dynamically linked library. 

\begin{example}
For instance, the {\em platform-independent version} of the Prot\'eg\'e tool (v5.2)\footnote{\url{https://protege.stanford.edu/}; last accessed on 16 July 2020.} is unzipped into three {\tt run} files for different operating systems and five folders with files, which include the core jar files such as {\tt protege-launcher.jar}, {\tt protege-editor- core.jar}, {\tt protege-editor-owl.jar}, and {\tt protege-common.jar}, as well as jar files for other essential features, such as {\tt owlapi-osgidistribution.jar}, and a bunch of plugins, such as the automated reasoner {\tt org.semanticweb.hermit-1.3.8.413.jar} and some query functionality {\tt org.coode.dlquery-4.0.1.jar}, further configuration settings (like in {\tt config.xml}) and the occasional icon for the interface. This is in contrast to the {\em platform-specific versions}, such as a {\tt Protege.exe}, which come as a single executable where all those separate files are bundled together and hidden from the user. That is, it is basically the same application, but it is offered in different modalities, which is entirely an engineering decision to cater for different user preferences, since the former, platform-independent version, is easier for, mainly, plugin management and less popular operating systems, whereas the latter is easier for end users to give it an out-of-the-box one-click-launch look-and-feel. 
\end{example}

Put differently, when there are separate files, at least some of them  could have been put `inside' the executable, but from an engineering and usability viewpoint, they allow for more flexible customisation of the program and support the prospect of reusability. In any case, regardless the variation in configuration, if there is more than one file for the computer program, they are {\em linked}---in the sense of the compilation processes in compilers, not the generic term `linked'---so that for compiled code, we obtain a similar definition as for the source code functional whole, but then $ linked(x_i,x_j,t)$ rather than with $SCgraphPath(x_i,x_j,t)$. There are several differences between the two relations: the latter is mutable and the former ($linked$) is not; the latter can have those files swapped, the former not; they hold between different types of files; and the latter has the links manually specified by the programmer's design, whereas this happens automatically in the compilation by the optimisation algorithms that are part of the compiler. 

Can $x_i,x_j$ be linked but not be part of a computer program $y$, or be genuine functional part but not linked, i.e., that the bi-implication does not hold? For code, there are {\em object code libraries} that are used in the linking stage of the compilation process, which $x_i,x_j$ are part of and while being part of that, they are not linked. With sub-optimal program code and that optional participation, surely also at least one example can be found where either $x_i$ or $x_j$ is not a $genuineFP$ of $y$. 
We thus obtain the following definition:

\begin{definition}[Compiled program as a functional whole]\label{def:cpdef}
For any computer program $C$ in compiled form that has one or more genuine functional parts $x_1, \ldots, x_n$ with $n \geq 1$ and $1 \leq i,j \leq n$, and 
time $t \in \mathcal{T}$ where $\mathcal{T}$ is the set of discrete time points, it holds that 
$\forall x_ix_jyt (genuineFP(x_i, y,t) \rightarrow (genuineFP(x_j,y,t) \leftrightarrow linked(x_i,x_j,t))) $.  For at least some $y$, $linked(x_i,x_j,t)$ does not hold, i.e.,: 
$\neg \forall x_ix_jt (linked(x_i,x_j,t) \leftrightarrow \exists y (genuineFP(x_i,y,t) \land genuineFP(x_j,y,t)))$. 
\end{definition}

\begin{figure}[t]
	\centering		
	\includegraphics[width=0.9\textwidth]{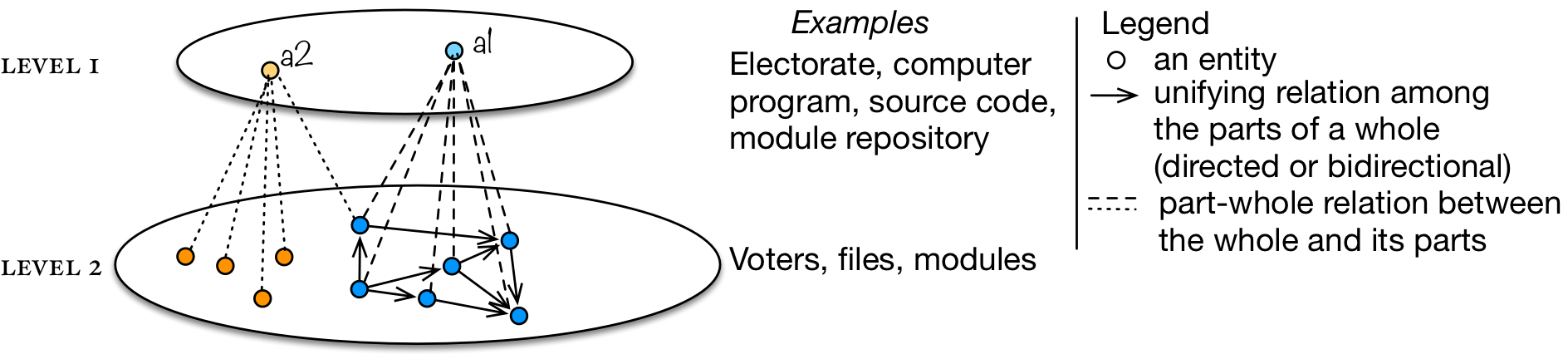}
	\vspace{-2mm}
		\caption{Levels of granularity, with entities residing in each level, which may be related `horizontally' among themselves at that level, and `vertically' between the parts and the whole each residing in a different level. The object {\sf a2} on the left visualises a collective with its members and {\sf a1} visualises an integral whole with its parts.} \label{fig:levels}
\end{figure}

The situation is sketched in Figure~\ref{fig:levels}, which shows two of the possibly many levels of granularity, each having entities residing at that level. The `vertical' relations between the whole at level $L_1$, and its parts at the finer-grained level $L_2$ are typically a part-whole relation, but may be different \cite{Keet08phd}. The relation among the parts at a particular level is the unifying relation, such as {\sf being in a source code graph with} and {\sf begin a fellow adult citizen} of a country. Informally, this constructs a triangle of relations between the entities: if you have one vertical side, there must be the other two (a horizontal and another vertical one).

\section{Computer programs as wholes, collections, or set?}
\label{sec:whole}

Having answered Q2 and Q3---$genuineFP$, $SCgraphPath$, and $linked$---we now can turn to Q1: is a program a whole, mere collection, or just a set? Clearly, it is not just a set of discrete artefacts, for the artefacts have a specific relation to each other, and they need to have that coupling to achieve the proper functioning of the program. Sets, on the other hand, do not pose such conditions on their members. Put differently: there are more things that are sets than that  are programs. For instance, one can put {\tt protege-launcher.jar} and {\tt pluraliser.py} together as a set, but that does not make them two parts of a computer program, since they are independent files because there is no `communication' between the two, notably not being in the same source code graph nor being linked, and not even more loosely connected through sockets or pipes. 

That program is not a collection either may be less obvious, and the arguments  depend on the definition of collection that one prefers. Let us consider each in turn. By Guizzardi's and Rector et al.'s requirement on the parts or grains to have to perform the same role \cite{Guizzardi05,Rector06}, a computer program is {\em not} a collective, because its components do not have all the same function or role nor are they necessarily the same type of file. Moreover, a crucial aspect of modularisation in software development is to separate  the different functions into different modules and files. Thus, when following established software design principles, it would  never be a collective by their characterisation. 

Considering Copp's criteria for collective \cite{Copp84}: the program can and does perform actions, therewith meeting one of the three criteria. Copp focusses on {\em social} collectives rather than any collective, and therefore also has the requirement that a collective is composed of persons and has a ``plausible theory of the legal system'', which clearly do not apply. He generalises collective so that possibly they are ``mereological sums of collective stages linked by a unity relation'' where his ``unity'' refers to a property of the whole to track the collective through time to determine diachronic identity and the mereological sum applies to aggregating the time-slices, not about the parts being summed into a whole. 
Diachronic identity is not applicable here as a criterion, supporting yet further the notion that software is not a collective. Thus, also by Copp's criteria and definition, a computer program is not a collective.

Tangentially, it does raise again the topic of mereological sum: could a software app be just a mereological sum? Bare mereological sums permit a sort of `contingent unity', however, like a sum of {\sf\em your right index finger and right thumb}. This is not what we have with program code, since each unity is {\em deliberate} by design and has properties at the level of the whole. Guarino and Welty tried to avoid simple mereological sums by assuming that  ``An object $x$ is an {\em intrinsic whole under} $\omega$ if, at any time where $x$ exists, it is a contingent whole under $\omega$.'' \cite{Guarino00a} where their $\omega$ is $B$ here. That is, the unifying relation $B$ is essential to the whole, whereas just a mereological sum need not to have $B$ holding among its parts all the time or even at some time. For instance, a Python module repository may contain the modules Owlready v0.3, Tkinter v5, 
and NLTK v2, which is a mereological sum and also a collective, yet the modules are independent from each other. The files in one's `Downloads' folder have even less unity: albeit sharing the characteristic of having been downloaded, each download, such as a {\tt yetAnotherVideoTool.pkg} and a {\tt 1234567.jpg}, is independent. The files of a particular version of the source code or computer program (without orphan files), however, {\em always} will have a binding relation to each other for as long as they exist; hence, then indeed is an intrinsic whole under $SCgraphPath$ or $linked$, respectively.  \\

Having used both the process of elimination and having provided supporting arguments as to what makes a whole a  whole (recall Definitions~\ref{def:scdef} and~\ref{def:cpdef}), a computer program---be it in source code form or as compiled code---qualifies as being a {\em whole}. It is a {\em functional} whole, since the function aspect comes from the whole having a function, which has `sub' or part-functions (recall Section~\ref{sec:fp}), as per design of the artefact. 
Thus, then the source code and the computer program are a {\em functional whole}. 

It perhaps invites one to propose a definition for what a computer program is. This, however, may at least in part be dependent on answering the ontological nature of a computer program as a whole \cite{Turner19}, which was not addressed in this paper---here it was argued only that it {\em is} a whole. For the situation in South Africa, where the author resides, there is no need for proposing a detailed definition, since it has been defined in its Copyright Act of 1978 that we have to adhere to. As noted in Section~\ref{sec:programnote}, ``computer program'' is  defined in the Act as: ``a set of instructions fixed or stored in any manner and which, when used directly or indirectly in a computer, directs its operation to bring about a result.''. Observe that the notion that the source code and compiled code are functional wholes is compatible with this definition, since both assert there is an entity (the program), the Act's definition does not make any statement about the files---be that component files of the program, in the sense of stored as human-readable source code, bytecode, or object code---so this cannot contradict, and the Act's statement is slightly stronger regarding function in the sense that the function actually has to work (cf. possibly not ever being realised).

\section{Conclusions}
\label{sec:concl}

An argumentation was presented that computer programs---be it as source code or as executable---is a functional whole, and why. In so doing, it had to draw from theories of mereology, granularity, unity, and function. 
The unifying relation among the parts (files) is the graph for source code and being linked for compiled code. The relation between the component files and the program is one of functional parthood, since the files perform (sub)functions of the overall function of the program. 
These additional insights into software and the notion of the internal structure of wholes may assist practically with, among others, litigation cases in software development, illegal downloads, and copyright infringements, as well as more generally by having demonstrated that the notion of a unifying relation is  indeed operationalisable.

An aspect that may be of interest for future work concerns the assumption of `free of bugs' and the modes of participation of the parts in  the whole: should one distinguish between those files that prevent a program from running altogether and those that cause one to lose some functionality but not its principal one and those where mainly just aesthetics are affected? That is, whether a sliding scale should be applied to determining importance of a source file to the whole program.\\

\noindent \textbf{Acknowledgements} I would like to thank the advocate and lawyer on the litigation case for the conversations on this topic, which assisted with devising examples and descriptions to try to explain the theory to a non-specialist audience. 


\begin{thebibliography}{10}

\bibitem{AGK08}
A.~Artale, N.~Guarino, and C.~M. Keet.
\newblock Formalising temporal constraints on part-whole relations.
\newblock In G.~Brewka and J.~Lang, editors, {\em 11th International Conference
  on Principles of Knowledge Representation and Reasoning (KR'08)}, pages
  673--683. AAAI Press, 2008.
\newblock Sydney, Australia, September 16-19, 2008.

\bibitem{Bjorner10}
D.~Bj{\o}rner and A.~Eir.
\newblock {\em Compositionality: Ontology and Mereology of Domains -- Some
  Clarifying Observations in the Context of Software Engineering}, pages
  22--59.
\newblock Springer Berlin Heidelberg, Berlin, Heidelberg, 2010.

\bibitem{Brachmann2018}
S.~Brachmann.
\newblock Arista pays cisco \$400m to end patent litigation at district court
  and itc.
\newblock {\em IPwatchdog}, 8 August 2018 2018.
\newblock Online:
  \url{https://www.ipwatchdog.com/2018/08/08/arista-pays-cisco-400m-end-patent-litigation-district-court-itc/id=100102/};
  Last accessed on 8 June 2020.

\bibitem{Cairns18}
P.~Cairns.
\newblock Does afrocentric have a r1bn problem?
\newblock {\em The Citizen}, 16 July 2018 2018.
\newblock Online:
  \url{https://citizen.co.za/business/1979904/does-afrocentric-have-a-r1bn-problem/};
  Last accessed on 8 June 2020.

\bibitem{Copp84}
D.~Copp.
\newblock What collectives are: Agency, individualism and legal theory.
\newblock {\em Dialogue}, 23:249–269, 1984.

\bibitem{Guarino00a}
N.~Guarino and C.~Welty.
\newblock Identity, unity, and individuality: towards a formal toolkit for
  ontological analysis.
\newblock In W.~Horn, editor, {\em Proceedings of the European Conference on
  Artificial Intelligence (ECAI'00)}, pages 219--223. IOS Press, Amsterdam,
  2000.

\bibitem{Guarino09oc}
N.~Guarino and C.~Welty.
\newblock An overview of ontoclean.
\newblock In S.~Staab and R.~Studer, editors, {\em Handbook on Ontologies},
  pages 201--220. Springer Verlag, 2009.

\bibitem{Guizzardi05}
G.~Guizzardi.
\newblock {\em Ontological Foundations for Structural Conceptual Models}.
\newblock Phd thesis, University of Twente, The Netherlands. Telematica
  Instituut Fundamental Research Series No. 15, 2005.

\bibitem{Karjiker16}
S.~Karjiker.
\newblock Copyright protection of computer programs.
\newblock {\em South African Law Journal}, 133(1):51--72, 2016.

\bibitem{Keet08phd}
C.~M. Keet.
\newblock {\em A Formal Theory of Granularity}.
\newblock Phd thesis, KRDB Research Centre, Faculty of Computer Science, Free
  University of Bozen-Bolzano, Italy, April 2008.

\bibitem{AK07}
C.~M. Keet and A.~Artale.
\newblock Representing and reasoning over a taxonomy of part-whole relations.
\newblock {\em Applied Ontology -- Special issue on Ontological Foundations for
  Conceptual Modeling}, 3(1-2):91--110, 2008.

\bibitem{KK18fois}
C.~M. Keet and L.~Khumalo.
\newblock On the ontology of part-whole relations in {Zulu} language and
  culture.
\newblock In S.~Borgo and P.~Hitzler, editors, {\em 10th International
  Conference on Formal Ontology in Information Systems 2018 (FOIS'18)}, volume
  306 of {\em FAIA}, pages 225--238. IOS Press, 2018.
\newblock 17-21 September, 2018, Cape Town, South Africa.

\bibitem{KXK17}
C.~M. Keet, M.~Xakaza, and L.~Khumalo.
\newblock Verbalising owl ontologies in isizulu with python.
\newblock In E.~Blomqvist, K.~Hose, H.~Paulheim, A.~Lawrynowicz, F.~Ciravegna,
  and O.~Hartig, editors, {\em The Semantic Web: {ESWC} 2017 Satellite Events},
  volume 10577 of {\em LNCS}, pages 59--64. Springer, 2017.
\newblock 30 May - 1 June 2017, Portoroz, Slovenia.

\bibitem{Koslicki13}
K.~Koslicki.
\newblock Substance, independence, and unity.
\newblock In E.~Feser, editor, {\em Aristotle on Method and Metaphysics},
  chapter~9, pages 169--195. Palgrave McMillan, 2013.

\bibitem{Le08}
D.~T.~M. L{\^e} and R.~Janicki.
\newblock {\em A Categorical Approach to Mereology and Its Application to
  Modelling Software Components}, pages 146--174.
\newblock Springer Berlin Heidelberg, Berlin, Heidelberg, 2008.

\bibitem{Mizoguchi08}
R.~Mizoguchi.
\newblock Functional ontology of artifacts.
\newblock In {\em Interdisciplinary Ontology Conference 2008}, pages 1--10.
  Keio University Press Inc., 2008.

\bibitem{Mizoguchi17}
R.~Mizoguchi and S.~Borgo.
\newblock A preliminary study of functional parts as roles.
\newblock In {\em FOUST-II: 2nd Workshop on Foundational Ontology, Joint
  Ontology Workshops 2017}, volume 2050 of {\em CEUR-WS}, page~9p, 2017.
\newblock 21-23 September 2017, Bolzano, Italy.

\bibitem{Mizoguchi16}
R.~Mizoguchi, Y.~Kitamura, and S.~Borgo.
\newblock A unifying definition for artifact and biological functions.
\newblock {\em Applied Ontology}, 11:129--154, 2016.

\bibitem{Noback18}
M.~Noback.
\newblock {\em Principles of Package Design: Creating Reusable Software
  Components}.
\newblock Apress, 1st ed. edition, 2018.

\bibitem{Prehofer08}
C.~Prehofer, J.~van Gurp, and J.~Bosch.
\newblock Compositionality in software product lines.
\newblock In A.~D. Lucia, F.~Ferrucci, G.~Tortora, and M.~Tucci, editors, {\em
  Emerging Methods, Technologies, and Process Management in Software
  Engineering}, chapter~2, pages 21--42. John Wiley \& Sons, 2008.

\bibitem{Rector06}
A.~Rector, J.~Rogers, and T.~Bittner.
\newblock Granularity, scale and collectivity: When size does and does not
  matter.
\newblock {\em Journal of Biomedical Informatics}, 39:333–349, 2006.

\bibitem{Samuelson19}
P.~Samuelson.
\newblock Api copyrights revisited.
\newblock {\em Communications of the ACM}, 62(7):20--22, 2019.

\bibitem{Tripakis16}
S.~Tripakis.
\newblock Compositionality in the science of system design.
\newblock {\em Proceedings of the IEEE}, 104(5):960--972, 2016.

\bibitem{Turner19}
R.~Turner, N.~Angius, and G.~Primiero.
\newblock The philosophy of computer science.
\newblock In E.~N. Zalta, editor, {\em The Stanford Encyclopedia of
  Philosophy}. Metaphysics Research Lab, Stanford University, spring 2019
  edition, 2019.

\bibitem{Varzi04}
A.~C. Varzi.
\newblock Mereology.
\newblock In E.~N. Zalta, editor, {\em Stanford Encyclopedia of Philosophy}.
  Stanford, fall 2004 edition, 2004.
\newblock \url{http://
  plato.stanford.edu/archives/fall2004/entries/mereology/}.

\bibitem{Vieu05}
L.~Vieu and M.~Aurnague.
\newblock Part-of relations, functionality and dependence.
\newblock In M.~Aurnague, M.~Hickmann, and L.~Vieu, editors, {\em
  Categorization of Spatial Entities in Language and Cognition}. John
  Benjamins, Amsterdam, 2005.

\bibitem{Winston87}
M.~Winston, R.~Chaffin, and D.~Herrmann.
\newblock A taxonomy of partwhole relations.
\newblock {\em Cognitive Science}, 11(4):417--444, 1987.

\end{thebibliography}

\end{document}